\documentclass{article}
\usepackage[T1]{fontenc} % add special characters (e.g., umlaute)
\usepackage[utf8]{inputenc} % set utf-8 as default input encoding
\usepackage{ismir,amsmath,cite,url}
\usepackage{graphicx}
\usepackage{color}
\title{Large Music Recommendation Studies for Small Teams}

\multauthor
{Kyle Robinson \hspace{1cm} Dan Brown} { \\
  David R. Cheriton School of Computer Science, University of Waterloo, Canada\\
{\tt\small kyle.robinson@uwaterloo.ca, dan.brown@uwaterloo.ca}
}
\def\authorname{Kyle Robinson, Dan Brown}

\def\authorname{K. Robinson, and D. Brown}

\usepackage[bookmarks=false,pdfauthor={\authorname},pdfsubject={\papersubject},hidelinks]{hyperref}
\begin{document}

\maketitle
\begin{abstract}
Running live music recommendation studies without direct industry partnerships can be a prohibitively daunting task, especially for small teams. In order to help future researchers interested in such evaluations, we present a number of struggles we faced in the process of generating our own such evaluation system alongside potential solutions. These problems span the topics of users, data, computation, and application architecture.
\end{abstract}

\section{Running a Live Recommendation Study}
There are clearly benefits to evaluating music recommender systems with real users~\cite{Knijnenburg2015}. In our recent paper analysing user perceptions of diversity in music recommendation we found that mere accuracy evaluations are not necessarily good indicators of individual track ratings, and overall list satisfaction is not a function of individual track ratings alone~\cite{Robinson2021}. Insights such as this are not new; a decade and a half ago McNee \textit{et al.} informally argued that there must be more emphasis put on user-centric recommender system evaluation~\cite{McNee2006}, yet just two years ago Dacrema \textit{et al.} highlighted a disturbing lack of attention to evaluation even in strictly offline analyses~\cite{FerrariDacrema2019}.

User studies and online analyses require significantly more resources and time than strictly offline analyses. In the hope of assisting researchers completing live evaluations of their methods, we present some of the struggles faced in developing our recent study, and their resolutions. For further reading on live evaluation of recommender systems we refer readers to the relevant chapters of the Recommender Systems Handbook~\cite{Gunawardana2015,Knijnenburg2015}.

\subsection{General Architecture}
The goals of our system were twofold: to generate up-to-date music recommendations for previously unseen participants using the same models described in pre-existing research, and to generate these recommendations on demand. The final system consisted of an online application which, after obtaining consent, collected participants' listening histories from the LastFM API, fed their data through several recommendation algorithms to obtain top-\textit{n} lists, obtained music previews and metadata from the Spotify API, and displayed song previews alongside song-specific appropriateness questions and global summary questions. We implemented the system using a Flask backend which served static HTML and Javascript\footnote{An un-maintained repository of our application can be found at https://github.com/Stack-Attack/music\_rec\_div\_study}. The study was hosted on a single AWS EC2 server instance using Elastic Beanstalk.

\subsection{Users and Data}

\subsubsection{Up-to-date Training Data}
A first challenge is that for collaborative filtering recommendation (especially for music) training data must be collected as close to the study as possible in order ensure that participant data is known by the model. Real data is also difficult to locate and obtain.

In the domain of music, LastFM continues to provide a useful API for user-song listening event (LE) data collection\footnote{The LastFM API documentation can be accessed at~\href{https://www.last.fm/api}{https://www.last.fm/api}}, though collection is not always a straightforward process. We collected a base data set from pseudo-randomly selected users to train our model by crawling the public social graph of friends lists. Contacting authors of prior research utilizing data sets which fit our needs proved to be vital in developing a method of data collection. Although their data was not up-to-date, we were able to develop our own collection method after corresponding. This data was topped-up before each subsequent study, and appropriately randomized.

\subsubsection{Live Participant Data}
An especially interesting challenge was that we needed to be able to collect data from participants as they connected to the system. This data also needed to be as current as possible.

Our solution to this problem was to use a media-tracking application. LastFM and its associated API also worked well for this purpose. For smaller pools of participants, we helped them register an account and manually monitored it over a collection period of a few weeks. For larger pools of participants we specified in recruitment and consent materials that they must have an existing account containing some minimum number of listening events/plays/scrobbles. Amazon Mechanical Turk specifically does not allow researchers to ask participants to log into any accounts, but because LastFM accounts are publicly accessible by default, data can be accessed with only a username. It is worth noting that in our case, some participants appear to have created new accounts to complete the study without being prompted to do so.

\subsubsection{Showing Music Recommendations}
To evaluate a recommendation, participants need to be able to listen to it!

Music previews can typically be accessed without having to authenticate with a music service. We used the Spotify API to obtain 30 second previews with album art in the form of HTML iframes embedded in the page~\footnote{Details on embedding Spotify music previews can be found at~\href{https://developer.spotify.com/documentation/widgets/guides/adding-a-spotify-embed/}{https://developer.spotify.com/documentation/widgets/guides/adding-a-spotify-embed/}}. The relevant track previews were retrieved by searching for tracks using their exact song and artist names as well as the region a participant was connecting form. We discarded the small portion of tracks that did not return any results; this is likely unavoidable.

\subsection{Computation}

\subsubsection{Model Training}
 We trained two different ML models for our project, one more traditional model based on matrix factorization, and one more modern approach based on neural networks. Writing the necessary code to implement models efficiently is time-consuming and error prone. Additionally, training models on huge data sets seems infeasible due to size and dimensionality.

Using open-source libraries can save time and help alleviate the risk of errors impacting results, though they may mis-implement key algorithmic features, or be difficult to extend. Comparing multiple implementations online can help, but one must ensure to follow any licenses and reference the source.

\subsubsection{Data Handling}
In order to reduce the size of data, we filtered out irrelevant items and users. By filtering out tracks with 10 or fewer LEs we reduced the number of unique tracks by 82\% while only decreasing LE count by 6\%. Even after filtering, our Variational Autoencoder (MultVAE) was too large to fit in GPU memory, and so we trained multiple model variations concurrently using CPU's in order to make up for lost time. Some models may simply be infeasible without access to High Performance Computing (HPC) resources. Other models may simply not be suited to real-world implementations without significant structural changes and/or pre-processing steps. This is simply the reality of evaluating models on real populations.

\subsubsection{Complex Architecture and Resource Requirements}
The size of trained models is too large to fit in memory, especially if a new model is loaded for each server connection.

The size of trained models can be \textbf{very} large even after removing unnecessary data (i.e., neural network optimizer information). Our trained MultVAE model was over 6.5GB in size.  Luckily, online cloud computing platforms often offer specific instances with large amounts of dedicated memory at the expense of processing power. These instances are a great fit for running user-studies which will inherently have a low number of concurrent users. As of now, simple hosting services such as Heroku will be infeasible due to the memory requirements~\cite{heroku}. AWS Elastic Beanstalk, however, provides very cost effective solutions in the form of memory optimized EC2 instances ~\cite{ec2}.

Even with the low number of concurrent users, there will still be some asynchronous computation required. The ideal, yet complex, solution to this problem is to decouple the longer tasks (recommendation and data collection) from the main server using a separate worker process or even server. Developing this architecture can be time-consuming, expensive, and unnecessary for such small temporary applications. We found success by limiting the server to one Python process, and running data collection and recommendation on separate threads using the built in \textit{concurrent.futures} library\footnote{In practice, we used the Flask-Executor python library to manage our futures: https://pypi.org/project/Flask-Executor/}. As live user data collection was input/output bound it did not block the server from handling requests, and as most recommendation and processing tasks utilized multi-core optimized libraries these tasks were also handled relatively quickly. This kind of architecture would certainly not work for large-scale applications, but was ideal for our small user-count study due to its simplicity and efficient use of only one compute node.

\subsection{Summary}
The benefits of evaluating music recommender systems on real users are as intuitive as they are founded in empirical evaluation~\cite{Knijnenburg2015}. Without industry collaboration, and at minimal cost, we were able to develop a music recommendation system which could generate and present recommendations to new users within a single un-moderated interactive session. To assist and encourage future researchers in developing similar systems, we have described some of the challenges and solutions to problems encountered along the way. Among the problems we addressed were training data collection, live user data collection, and obtaining music previews. We also discussed our technical implementation; specifically dealing with issues of memory management and availability. In general, we hope that researchers embrace collaboration with others to better base our analysis of recommender systems in users themselves. Our key message is that independent user studies on music recommendation are both important and achievable.

% For bibtex users:
\bibliography{ISMIRtemplate}

% For non bibtex users:
%\begin{thebibliography}{citations}
% \bibitem{Author:17}
% E.~Author and B.~Authour, ``The title of the conference paper,'' in {\em Proc.
% of the Int. Society for Music Information Retrieval Conf.}, (Suzhou, China),
% pp.~111--117, 2017.
%
% \bibitem{Someone:10}
% A.~Someone, B.~Someone, and C.~Someone, ``The title of the journal paper,''
%  {\em Journal of New Music Research}, vol.~A, pp.~111--222, September 2010.
%
% \bibitem{Person:20}
% O.~Person, {\em Title of the Book}.
% \newblock Montr\'{e}al, Canada: McGill-Queen's University Press, 2021.
%
% \bibitem{Person:09}
% F.~Person and S.~Person, ``Title of a chapter this book,'' in {\em A Book
% Containing Delightful Chapters} (A.~G. Editor, ed.), pp.~58--102, Tokyo,
% Japan: The Publisher, 2009.
%
%
%\end{thebibliography}

\end{document}